\documentclass[conference]{IEEEtran}
\IEEEoverridecommandlockouts
\usepackage{booktabs}
\usepackage{graphicx}
\usepackage{quantikz}
\usepackage{braket}
\usepackage{wrapfig}
\usepackage{cite}
\usepackage{amsmath,amssymb,amsfonts}
\usepackage{algorithmic}
\usepackage{graphicx}
\usepackage{textcomp}
\usepackage{xcolor}
\def\BibTeX{{\rm B\kern-.05em{\sc i\kern-.025em b}\kern-.08em
    T\kern-.1667em\lower.7ex\hbox{E}\kern-.125emX}}
\begin{document}

\pgfset{
    framed/.style={
        /tikz/commutative diagrams/every label/.append style={
            fill=white,
            rounded corners=2.5pt,
            font={\normalsize},
            anchor=center,
            #1
        }
    },
    framed/.default={draw=red}
}    

\title{QCA-MolGAN: Quantum Circuit Associative Molecular GAN with Multi-Agent Reinforcement Learning\\

\thanks{Funded by the European Union, Project FULLMAP number 101192848}
}

\author{\IEEEauthorblockN{1\textsuperscript{st} Aaron Mark Thomas}
\IEEEauthorblockA{\textit{Department of Computer Science} \\
\textit{University of Birmingham}\\
Birmingham, United Kingdom, \\
\textit{QunaSys Europe}\\
Copenhagen, Denmark\\
amt326@student.bham.ac.uk}
\and
\IEEEauthorblockN{2\textsuperscript{nd} Yu-Cheng Chen}
\IEEEauthorblockA{\textit{Hon Hai Research Institute} \\
Taipei, Taiwan \\
kesson.yc.chen@foxconn.com}
\and
\IEEEauthorblockN{3\textsuperscript{rd} Hubert Okadome Valencia}
\IEEEauthorblockA{\textit{QunaSys Inc} \\
Tokyo, Japan \\
hubert@qunasys.com}
\and[\hfill\mbox{}\par\mbox{}\hfill]
\IEEEauthorblockN{4\textsuperscript{th} Sharu Theresa Jose}
\IEEEauthorblockA{\textit{Departement of Computer Science} \\
\textit{University of Birmingham}\\
Birmingham, United Kingdom \\
s.t.jose@bham.ac.uk}
\and
\IEEEauthorblockN{5\textsuperscript{th} Ronin Wu}
\IEEEauthorblockA{\textit{QunaSys Europe} \\
Copenhagen, Denmark \\
ronin@qunasys.com}
}

\maketitle

\begin{abstract}
Navigating the vast chemical space of molecular structures to design novel drug molecules with desired target properties remains a central challenge in drug discovery. Recent advances in generative models offer promising solutions. This work presents a novel quantum circuit Born machine (QCBM)-enabled Generative Adversarial Network (GAN), called QCA-MolGAN, for generating drug-like molecules. The QCBM serves as a learnable prior distribution, which is associatively trained to define a latent space aligning with high-level features captured by the GANs discriminator. Additionally, we integrate a novel multi-agent reinforcement learning network to guide molecular generation with desired targeted properties, optimising key metrics such as quantitative estimate of drug-likeness (QED), octanol-water partition coefficient (LogP) and synthetic accessibility (SA) scores in conjunction with one another. Experimental results demonstrate that our approach enhances the property alignment of generated molecules with the multi-agent reinforcement learning agents effectively balancing chemical properties.
\end{abstract}


\section{Introduction}

Designing novel molecules with targeted properties for drug and material development is a challenging task which has attracted significant attention \cite{meanwell2011improving, popova2018deep,zunger2018inverse}. The number of possible synthesizable molecules in the vast chemical space is estimated to be of the order $10^{23}-10^{100}$ drug compounds \cite{van2019virtual}. Traditionally, identifying lead candidate molecules has been a lengthy and costly part of the drug discovery pipeline \cite{hughes2011principles}. To address this challenge, data-driven machine learning methodologies present a possible automated approach to accelerate this process \cite{vamathevan2019applications, dara2022machine}.

Deep generative models have gained lots of traction in drug discovery research, leveraging various frameworks for molecular data representations. This includes graph-based representations with node and edge attributes corresponding to atoms and bonds \cite{goodfellow2014generative, kingma2013auto, medsker2001recurrent}, or the SMILES representation, a string-based representation of a molecule \cite{weininger1988smiles}. These frameworks leverage the concept of Quantitative Structure-Activity Relationships (QSAR) \cite{perkins2003quantitative}, where the structural attributes of molecules correlate directly with their chemical and biological properties. By learning an inverse mapping of this relationship, generative models enable the design of molecular structures guided explicitly by targeted properties. Such models include the Generative Adversarial Networks (GANs) \cite{goodfellow2014generative}, Variational Autoencoders (VAEs) \cite{kingma2013auto} and Recurrent Neural Network (RNNs) architectures \cite{medsker2001recurrent}. Recently, likelihood-free GAN architectures, such as MolGAN \cite{de2018molgan}, have been successfully applied in targeted molecular design. 

Quantum computing is a new computational paradigm that leverages quantum resources to efficiently sample probabilistic distributions considered intractable for classical computers \cite{shepherd2009temporally}. Recent advancements in quantum computing have triggered the development of quantum generative models that can potentially capture complex correlations in the data to efficiently generate new data. However, due to the constraints imposed by the current era of Noisy Intermediate-Scale Quantum (NISQ) computers \cite{preskill2018quantum}, recent proposals leverage the Hybrid Quantum Computing (HQC) framework where quantum subroutines are used along with classical algorithms to accelerate their performance \cite{biamonte2017quantum, madsen2022quantum}. Recent studies have explored the use of HQC schemes for molecular design \cite{li2021quantum, gircha2023hybrid,kao2023exploring, anoshin2024hybrid}. This includes QGAN-H, a hybrid quantum-classical generator within a GAN framework \cite{li2021quantum}, and more recently, Hybrid Quantum Cycle MolGAN \cite{anoshin2024hybrid},  which incorporates a cycle component into the architecture to achieve further improvement in small drug discovery. Despite the potential of such hybrid quantum generative frameworks, they suffer from several fundamental challenges: (a) scalability of quantum models for real-world classical datasets \cite{wossnig2021quantum}; (b) the issue of mode collapse in GAN based architectures, including MolGAN, where the generated data fails to capture all the modes of the underlying data distribution \cite{tomar2023review}.

\begin{figure}[!t]
    \centering
    \includegraphics[width=0.9\linewidth]{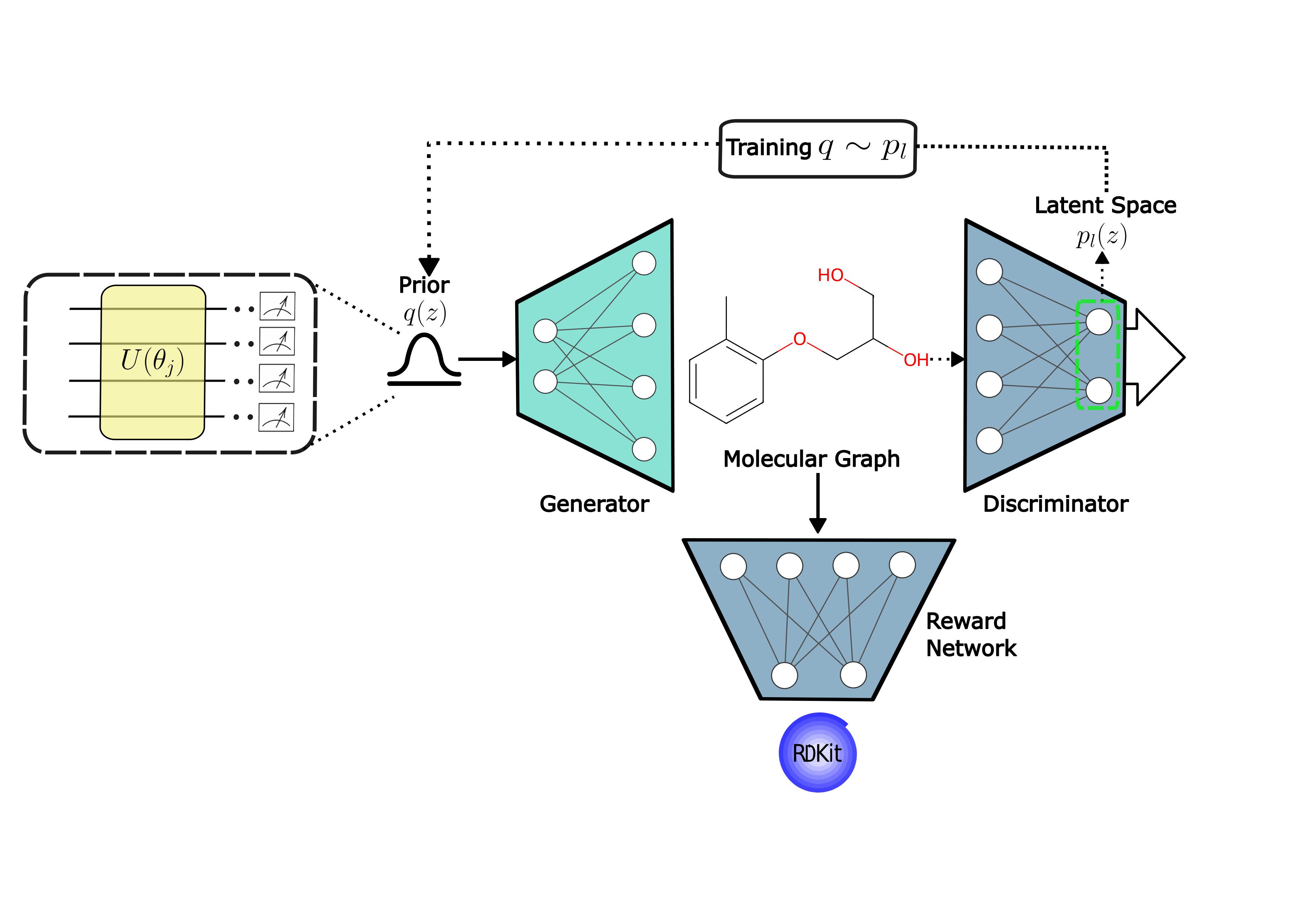}
    \vspace{-1.2cm}
    \caption{\textbf{Pipeline of Associative adversarial quantum MolGAN with multi-agent RL.} We first sample bitstring samples from the computational basis of a quantum circuit born machine, which acts as the GAN's prior distribution. The generator then generates candidate molecule graphs, guided by the multi-agent RL network to maximise several target properties. At the end of an epoch, the prior is trained to match a bottleneck layer of the discriminator.}
    \label{fig:pipeline}
\end{figure}

To address these shortcomings, this paper introduces a Quantum Circuit Associative MolGAN (QCA-MOLGAN) framework for generating graph representations of small molecules. The proposed QCA-MolGAN framework uses a quantum circuit born machine (QCBM) as a probabilistic generative model to provide latent samples to the classical GAN model. Specifically,  the QCBM describes a discrete prior distribution over the space of $n$-length bit-strings $\{0,1\}^n$, when using $n$-qubits. Importantly, the QCBM is trained to match the output distribution of the activation nodes of a deep latent layer of the discriminator, providing a high-level associative memory to the MolGAN framework during training. The resulting QCA-MolGAN model is highly amenable to run on current NISQ devices as the quantum and classical optimisation schemes are disentangled by separately sampling from a quantum device/simulator. 

\textbf{Our main contributions are as follows:}
\begin{enumerate}
    \item We propose a novel hybrid quantum-classical generative architecture, QCA-MolGAN, that integrates a QCBM as an associative adversarial network component. The trained QCBM prior significantly enhances the generative diversity of the typical MolGAN architecture. 
    \item We incorporate Multi-Agent Reinforcement Learning (MARL) into the generative pipeline, employing specialized agents to individually predict distinct molecular properties. These agents then collaboratively guide molecular generation resulting in optimised molecule design across multiple metrics simultaneously. Our experiments on QM9 5k dataset conclusively shows the superior macro average across all the key properties.
    
\end{enumerate}

\section{Preliminaries}\label{sec:preliminaries}

\noindent
\textbf{Molecular Graph Representation.}  
In line with previous works \cite{de2018molgan, tsujimoto2021molgan}, we represent a molecule as a graph \( G = (A, X) \), where \( A \) denotes the adjacency tensor and \( X \) denotes the node feature matrix. For a graph with \( N \) nodes and \( d \) distinct node types, the node feature matrix \( X \in \{0,1\}^{N \times d} \), with each row representing a one-hot encoding of the node type. The adjacency tensor \( A \in \{0,1\}^{N \times N \times b} \) encodes the presence of edges between nodes, with \( b \) denoting the number of edge (bond) types. Specifically, there is a bond of type \( k \) between nodes \( i \) and \( j \) if \( A_{ijk} = 1 \). Following the MolGAN paper, we encode the graph representation $G$ of a molecule via a relational graph neural network, that can capture structural relationships within molecular graphs \cite{wu2020comprehensive, zhou2020graph, schlichtkrull2018modeling}. For notational convenience, in the following sections, we use $x$ to denote the graph $G$ of a molecule.

\noindent
\textbf{Generative Adversarial Networks.} Generative Adversarial Networks (GANs) \cite{goodfellow2014generative} are a type of implicit generative learning model that attempts to learn the probability distribution underlying an observed data set by playing a 2-person zero-sum game \cite{mohebbi2023games}. A GAN consists of a pair of differentiable functions representing the generator, $G_{\phi}$, and the discriminator $D_{\omega}$, typically modelled by neural networks parameterised by vectors $\phi$ and $\omega$ respectively. The generator $G_{\phi}(z)=\tilde{x}$ maps a  latent input vector $z$ to a synthetic molecular graph $\tilde{x}$, while the discriminator $D_{\omega}(x)$ (or $D_{\omega}(\tilde{x}$)) assigns a discriminating score to the true graph $x$ (or the generated graph $\tilde{x}$). The GAN optimisation process pits the generator and discriminator neural networks against each other in an adversarial game.

\noindent
\textbf{Reinforcement Learning.} Reinforcement learning (RL) is an online learning paradigm where an agent learns to take an action by optimising an action policy through interactions with an environment \cite{ladosz2022exploration}. One such policy is the deep deterministic policy gradient where the agent attempts to take actions that maximise an approximation of a future expected reward \cite{silver2014deterministic, lillicrap2015continuous}. 

RL has been recently used in graph molecular generation \cite{popova2018deep}, where it guides the generation of new molecules with desirable performance metrics. Note that for each molecular graph $G$, the structure-activity relationship with respect to a chemical property of interest can be quantified via several metrics. This includes the Quantitative Estimate of QED score \cite{bickerton2012quantifying}, LogP score \cite{comer2001lipophilicity}, and the SA score \cite{ertl2009estimation}. 

Mathematically, let $f(\cdot)$ denote the structure-activity relationship function that assigns a score or ``reward" $f(G)$ to each input graph $G$. We use a $\beta$ parameterised neural network function $f_{\beta}(\cdot)$ to approximate the unknown reward function $f(\cdot)$, yielding a differentiable model amenable to training. 
We then use the reward feedback signal in conjunction with the GAN  to bias the exploration of the molecular space to specific regions of chemical interest. More details can be found in the next section.

\section{Methodology}

This section presents the proposed QCA-MolGAN generative model for small drug-like molecular generation. As can be seen in Figure~\ref{fig:pipeline}, QCA-MolGAN consists of three key sub-structures: the associative adversarial framework consisting of a quantum circuit-based prior latent distribution, the generator and discriminator of the GAN architecture, and the multiple chemical property prediction framework crucial to generating molecules that satisfy multiple properties. We detail each of the substructures and the training scheme in the following sub-sections.

\subsection{Quantum Circuit-Based Associative Adversarial Networks}

Associative Adversarial Networks (AANs) \cite{arici2016associative} were introduced as an extension of the traditional GAN to incorporate an associative memory component into the architecture. Such networks can effectively model prior latent space distributions
by learning high-level features from the bottleneck layer of the discriminator, thereby 
enhancing the generative capabilities of the network \cite{ackley1985learning, wilson2021quantum,rudolph2022generation}.
Motivated by recent work \cite{rudolph2022generation}, this paper considers a quantum circuit Born machine (QCBM) as a trainable prior distribution $q_{\theta}(z)$ on the latent space $\mathcal{Z}$. 

A QCBM is defined by a parameterised quantum circuit that applies a unitary operator $U(\theta)$ on an initial $n$-qubit  ground state $\vert 0 \rangle ^{\otimes n}$ to get the quantum state
\begin{equation}
    \ket{\Psi(\theta)} = U(\theta)\ket{0}^{\otimes n}.
\end{equation} The probability of observing any of the $2^n$ bit strings $z$ is then given by Born's rule as
\begin{equation}
q_{\theta}(z)
= \bigl|\langle z \mid \Psi(
\theta) \rangle\bigr|^2 \,.
\end{equation} Thus, the QCBM describes a distribution $q_{\theta}(z)$ on the discrete latent space $\mathcal{Z}$ of $2^n$ possible bit strings.

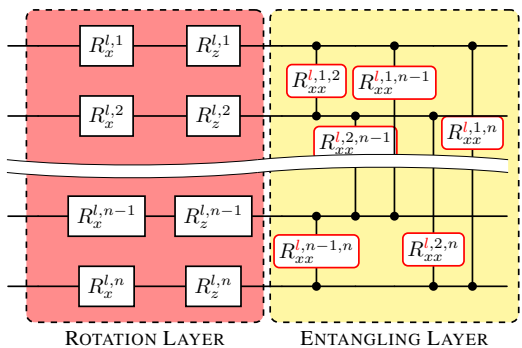
\begin{figure}
\begin{center}
\begin{tikzpicture}
\node[scale=0.8] {
 \begin{quantikz}\label{circuit_diagram}
&\gategroup[5,steps=3,style={dashed,rounded corners,fill=red!45, inner xsep=2pt},background,label style={label position=below,anchor=north,yshift=-0.2cm}]{{\sc Rotation Layer}}&\gate{R_x^{l,1}} & \gate{R_z^{l,1}} 

&\gategroup[5,steps=7,style={dashed,rounded corners,fill=yellow!45, inner xsep=2pt},background,label style={label position=below,anchor=north,yshift=-0.2cm}]{{\sc Entangling Layer}}

&\ctrl[style={framed}, 
    wire style={"R^{\textcolor{red}{l},1,2}_{xx}"}]{1}  &        & \ctrl[style={framed}, 
    wire style={"R^{\textcolor{red}{l},1,n-1}_{xx}"}]{1}       & &\ctrl[style={framed}, 
    wire style={"R^{\textcolor{red}{l},1,n}_{xx}"}]{3} & \\

&&\gate{R_x^{l,2}} & \gate{R_z^{l,2}} & &\ctrl{0} &\ctrl[style={framed}, 
    wire style={"R^{\textcolor{red}{l},2,n-1}_{xx}"}]{1}&        &    \ctrl{2}   & &  \\
\wave&&&&&&&&&&& \\
&&\gate{R_x^{l,n-1}} & \gate{R_z^{l,n-1}} &  & \ctrl[style={framed}, 
    wire style={"R^{\textcolor{red}{l},n-1,n}_{xx}"}]{1}&\ctrl{-1} &\ctrl{-2}  &       & & \\

 &&\gate{R_x^{l,n}} & \gate{R_z^{l,n}} &        &    \ctrl{0}    &        & &\ctrl[style={framed}, 
    wire style={"R^{\textcolor{red}{l},2,n}_{xx}"}]{-1}& \ctrl{-1}&
\end{quantikz}};
\end{tikzpicture}
\caption{QCBM: Unitary learning layer composed of a rotation layer with $R_x$ and $R_z$ parametrised gates, followed by all-to-all entangling operations using controlled $R_{xx}$ gates.
} 
\label{fig:quantum_subgenerator}
\end{center}
\end{figure}

We consider the QCBM architecture as shown in Figure~\ref{fig:quantum_subgenerator}, with an $L$-layer PQC that implements the unitary $U(\theta)$ as  
\begin{equation}
    U(\theta) = \prod_{l=1}^L \left(U_{\text{ent}}(\theta^l_\text{ent}) U_s(\theta_s^l)\right), 
\end{equation} where each $l$th layer consists of a rotation unitary gate $U_s(\theta_s^l)$, followed by an entangling unitary gate $U_{\text{ent}}(\theta^l_\text{ent})$. The rotation unitary gate consists of single-qubit rotations implemented as \begin{equation}
    U_s(\theta_s^l) = \bigotimes_{i=1}^{n}R_z(\theta_z^{l,i})R_x(\theta_x^{l,i}),
\end{equation}
where $R_x(\theta^{l,i}) 
= e^{-i\frac{\theta^{l,i}}{2}\hat{\sigma}_x}$ and $R_z(\theta^{l,i}) 
= e^{-i\frac{\theta^{l,i}}{2}\hat{\sigma}_z}$, with $\hat{\sigma}_x$ and $\hat{\sigma}_z$ denoting the Pauli X and Z matrices, represent the rotations around $X$ and $Z$ axes, respectively. The angles $\theta_x^{l,i}$ (or $\theta_z^{l,i}$) denote the rotation angle of the $l$th-layer gate acting on $i$th qubit.  We use this notation interchangeably with $R_x^{l,i} \equiv R_x(\theta^{l,i})$. Additionally, the $l^{th}$ entangling unitary is implemented as
\begin{equation}
    U_{ent}(\theta^l_\text{ent}) = \prod_{1\leq i <j\leq n} R_{xx}(\theta_{xx}^{l,i,j}),
\end{equation}
where $R_{xx}(\theta^{l,i,j}) = e^{-i\frac{\theta^{l,i,j}}{2}\hat{\sigma}^i_x\hat{\sigma}^j_x}$  denotes the two-qubit gate in the $l$th layer, that is applied between qubits $i$ and $j$. Note that the above two qubit-gate is applied across all pairs of qubits, creating an all-to-all entangling. 

In associative adversarial training, the QCBM parameters are tuned to match the probability distribution of the node activations in the $l$th deep latent layer of the discriminator with matching size $N$ \cite{wilson2021quantum, urushibata2022comparing, rudolph2022generation}. The $l$th layer captures condensed feature representation from the true and generated data forming a high level associative memory. This is achieved by training the QCBM parameters to minimize the cross-entropy loss with respect to the distribution $p_l(z)$ of the $l$-th feature layer of the discriminator via the following training criterion,
\begin{align}
    \mathcal{L}_{\text{AAN}} &= \max_{\theta} \mathbb{E}_{z \sim p_l(z)} \left[ \log q_\theta(z) \right]. \label{eq:AAN}
\end{align} In \eqref{eq:AAN}, the expectation is over samples $z$ drawn from the $l$th discriminator layer distribution. In practice, we use a clipped log-likelihood $\max \{\log q_{\theta}(z),\epsilon \}$ in \eqref{eq:AAN},  with some clipping constant $\epsilon>0$, which prevents collapse of the logarithm evaluation when encountering a sample $z$ with $q_{\theta}(z)=0$.

\subsection{GAN framework}
We now introduce the GAN architecture of our proposed QCA-MolGAN model. While several proposals for GAN implementations exist, we use Wasserstein GANs, which employ the Wasserstein distance between true and generated data as the optimization objective, owing to its improved training stability and better resilience to mode collapse. Specifically, we consider a practical implementation of Wasserstein GAN proposed in \cite{gulrajani2017improved}, which imposes a
gradient penalty on the discriminator to ensure that the family of parameterised discriminator functions are 1-Lipschitz continuous. 

Formally, let $G_\phi$ and $D_\omega$ represent the generator and discriminator (or critic) of our QCA-MolGAN with parameters $\phi$ and $\omega$ respectively. Further, let $x=G=(A,X)$ and $\tilde{x}= \tilde{G}=(\tilde{A},\tilde{X})$ denote the  true and generated molecular graph samples, respectively.   Following \cite{de2018molgan}, the min-max optimisation problem for the WGAN is then defined as:

\begin{align}
\mathcal{L}_{\text{WGAN}}(\phi,\omega) =\ & 
\mathbb{E}_{x \sim p_{r}(x)} \left[ D_{\omega}(x) \right] 
- \mathbb{E}_{z \sim q_{\theta}(z)} \left[ D_{\omega}(G_{\phi}(z)) \right] \notag \\
& + \lambda \mathbb{E}_{\hat{x} \sim \hat{p}(x)} \left[ 
\left( \left\| \nabla_{\hat{x}} D_\omega(\hat{x}) \right\|_2 - 1 \right)^2 
\right]
\end{align}
where $p_r(x)$ denote the true data distribution,  $q_{\theta}(z)$ is the QCBM prior distribution,  $\lambda$ is the penalty coefficient and $\hat{p}(x) = \epsilon p_r(x) + (1 - \epsilon) p_{g}(x)$ denote the mixture of the true distribution $p_r(x)$ and generated distribution $p_{g}(x)$, with $\epsilon$ sampled from a uniform distribution. Here, $p_g(x)$ is the generated distribution obtained via $z \sim q_{\theta}(z)$ and $\tilde{x}=G_{\phi}(z)$.

We consider a a multilayer perceptron (MLP) network as the generator that maps the latent input $z$ into a synthetic output graph $\tilde{x}= G_{\phi}(z)$ in a one-shot approach. Note that the resulting fake sample $\tilde{x}$ corresponds to the dense node and adjacency tensors $\tilde{x} = (\tilde{A},\tilde{X})$. To recover the molecular graph from the node and adjacency tensors, we first 
apply a softmax operation ($\text{softmax}(x)_i = \exp(x_i) / \sum_{i=1}\exp(x_i)$) along the last dimension to normalize the outputs into valid probability distributions. We then apply an argmax operation over the last dimension of $\tilde{X}$ and $\tilde{A}$ to obtain the most probable atom and bond types at each node and edge, yielding the molecular graph. In contrast, the discriminator network uses a series of graph convolutions and aggregation methods to process the true and synthetic graphs.

\label{sec:marl}
\subsection{Multi-Agent Reinforcement Learning}

In the reinforcement learning framework, the generator $G_{\phi}(\cdot)$ acts as the policy network that maps prior latent samples $z$ to molecular graphs (or `actions') $(\tilde{A}, \tilde{X}) =G_{\phi}(z)$. As discussed in Section~\ref{sec:preliminaries}, we then use a tunable reward-prediction function $f_{\beta}(\cdot)$  to estimate the reward associated with the generated graph. The reward network requires scalar rewards $\in (0,1)$ and the generator is subsequently optimised to maximise this prediction, providing gradients which guide molecular generation toward a desirable metric.

To facilitate multi-property optimisation, we propose a MARL framework that incorporates $M$ pretrained chemical property prediction networks which serve as specialised reinforcement learning agents \cite{tang2024molecular}. Each agent independently estimates and optimises a distinct molecular property, including QED, LogP, and SA scores. The architecture of each agent is identical to that of the discriminator, composed of graph processing units as well as dense layers to estimate the chemical property score.

Specifically, for the $i$th molecular property, $i=1,\hdots,M$, the true property value $r_i(\tilde{x}) \in \mathbb{R}$ of molecule $\tilde{x}$ is calculated using RDKit. The $i$th agent is trained independently to predict and optimise a specific molecular property $f_i(\tilde{x};\beta_i)$, such that it minimises the following loss objective: 
\begin{equation}
    \mathcal{L}_i = \mathbb{E}_{z \sim q_{\theta}(z)} [(f_i(G_{\phi}(z); \beta_i) - r_i(G_{\phi}(z)))^2].
\end{equation} The loss function above ensures that the estimated molecular property well-approximates the true property $r_i(\tilde{x})$ calculated.

\begin{figure}
    \centering
    \includegraphics[width=1\linewidth]{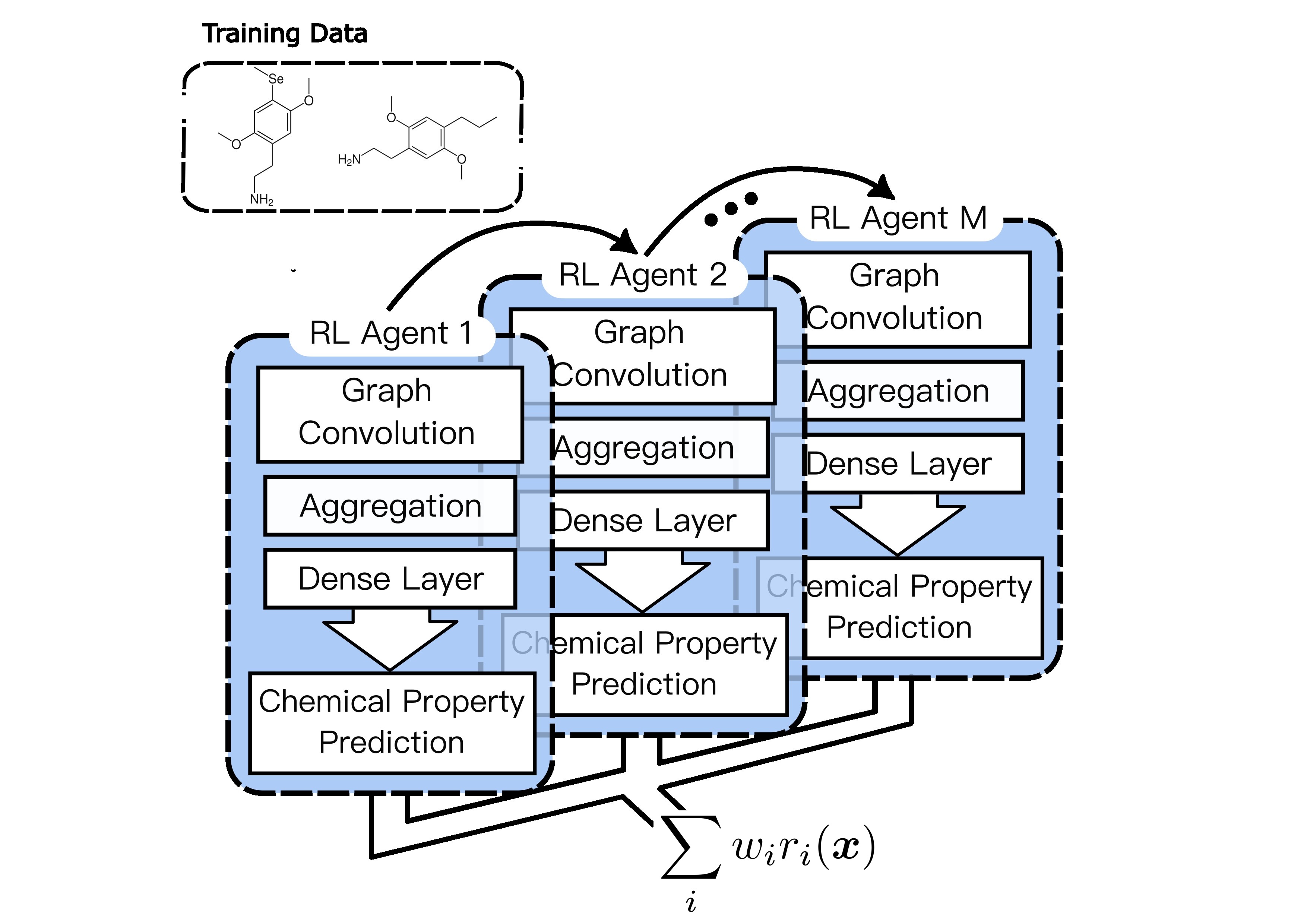}
    \caption{Multi Agent Reinforcement Learning Network: A series of RL agents are individually trained to each predict a given chemical property of a molecule. Once fully trained, a weighted sum of their predictions is maximised as the policy of the generator network.}
\end{figure}

The generator is trained to maximise the following expected aggregate reward from the agents,
\begin{equation}
\begin{aligned}
  \mathcal{L}_{\rm MARL}
  &= \mathbb{E}_{z \sim q_{\theta}(z)}\bigl[R(G_\phi(z))\bigr] \\
  &= \mathbb{E}_{z \sim q_{\theta}(z)}\Bigl[\sum_{i=1}^M w_i\,f_i\bigl(G_\phi(z);\beta_i\bigr)\Bigr] 
\end{aligned},
\end{equation}
where 

$R(x) = \sum_i w_i f_i(x;\beta_i)$  denotes the aggregated reward, and  $w_i$ denotes the weighting factor  representing the relative importance of each property, satisfying $\sum_{i=1}^M w_i = 1$. 
%
This multi-agent approach thus allows the generation process to simultaneously optimize multiple molecular objectives, leading to a richer and more versatile exploration of chemical space.

\subsection{Objective Function}

Our QCA-MolGAN model integrates three distinct training objectives to form a unified learning framework, leveraging the strengths of a quantum associative memory, Wasserstein GANs, and multi-agent reinforcement learning. Each component targets specific aspects of molecular graph generation. The combined loss function that encapsulates these three aspects is defined as follows:
\begin{equation}
\mathcal{L} = \mathcal{L}_{\text{AAN}} + \gamma\mathcal{L}_{\text{WGAN}} + (1 - \gamma) \mathcal{L}_{\text{MARL}}
\end{equation}
where $\gamma$ controls the relative strengths of the reinforcement learning and WGAN components. Initially, the model is trained without the inclusion of $\mathcal{L}_{\text{MARL}}$ into the loss as pretraining of these networks is critical to ensure accurate prediction of molecular properties.

\section{Experiments}

\begin{table*}[t]
\centering
\footnotesize
\resizebox{\linewidth}{!}{
  \begin{tabular}{lcccccccc}
    \toprule
    \textbf{Prior (Metric)} & \textbf{Valid (\%)} & \textbf{Unique (\%)} & \textbf{Novel (\%)} & \textbf{Diversity} & \textbf{Druglikeness (QED) } & \textbf{Synthesizability (SA) } & \textbf{Solubility (LogP) } & \textbf{Average}\\
    \midrule

    QCBM (QED)  & 99.7 $\pm$ 0.4  & \textbf{31.3} $\pm$ \textbf{18.3} & 99.4 $\pm$ 1.0  & 0.873 $\pm$ 0.122 & \textbf{0.607} $\pm$ \textbf{0.015} & 0.578 $\pm$ 0.065 & 0.437 $\pm$ 0.039 & 0.541 \\
    QCBM (LogP) & 99.9 $\pm$ 0.1  & 16.6 $\pm$ 16.6                 & \textbf{100.0 $\pm$ 0.0} & 0.947 $\pm$ 0.077 & 0.459 $\pm$ 0.069 & 0.802 $\pm$ 0.118 & 0.685 $\pm$ 0.013 & 0.649 \\
    QCBM (SA)   & \textbf{100.0} $\pm$ \textbf{0.0} & 6.9 $\pm$ 2.2      & \textbf{100.0} $\pm$ \textbf{0.0} & 0.980 $\pm$ 0.013 & 0.498 $\pm$ 0.078 & 0.946 $\pm$ 0.064 & 0.582 $\pm$ 0.123 & 0.675 \\
    QCBM (MARL) & \textbf{100.0} $\pm$ \textbf{0.0} & 4.5 $\pm$ 0.6      & \textbf{100.0} $\pm$ \textbf{0.0} & \textbf{1.000} $\pm$ \textbf{0.001} & 0.498 $\pm$ 0.013 & \textbf{0.958} $\pm$ \textbf{0.051} & \textbf{0.706} $\pm$ \textbf{0.008} & \textbf{0.721} \\
    \bottomrule
  \end{tabular}
}

\ 

\caption{Evaluation metrics using different priors (optimised metric in parentheses). Bold indicates the best (highest mean) per column.}

\label{tab:qcbm_metrics}
\end{table*}

\vspace{0.5em}

\subsection{Experimental Setup}
We conduct experiments using PyTorch/PyTorch Lightning for the classical generative models and PennyLane for the quantum circuit simulation environment. We run all experiments on an Apple MacBook Air (M2, 8~GB RAM)

\noindent
\textbf{Datasets:}  We validate QCA-MolGAN on a subset of the widely used QM9 dataset for small drug discovery \cite{ramakrishnan2014quantum, ruddigkeit2012enumeration}. The dataset comprises approximately 5000 neutral molecules, each with no more than nine heavy atoms (Carbon (C), Oxygen (O), Nitrogen (N), Fluorine(F)).

\ 

\noindent
\textbf{Chemical Properties and Metrics:}
In our experiments, we assess the performance of our model and generated molecules along two complementary axes: 

\begin{itemize}
    \item \textbf{Chemical Properties}: We compute standard chemical property descriptors using RDKit. This includes QED, which captures how drug-like a molecule is using several descriptors, LogP, which quantifies a molecule’s hydrophilic and lipophilic balance, and SA score, which captures the pharmacological relevance and computes the synthetic feasibility of compounds. 
    
    \item \textbf{Model Metrics}: We evaluate our model using four key metrics: \emph{validity} – the fraction of molecules that are generated that conform to chemical valence rules; \emph{uniqueness} – the proportion of valid molecules that are distinct from one another; \emph{novelty} – the share of unique molecules not found in the training set; and \emph{diversity} – a measure of structural variability among the generated molecules.
\end{itemize}

We train QCA-MolGAN for 300 epochs with a batch size of 32. We set a learning rate of $1\times10^{-3}$ for all components (generator, discriminator, RL). We pre-train the RL module for 150 epochs using $\mathcal{L}_{\text{MARL}}$ and the QM9 5K data before then integrating it into the generator loss $\mathcal{L}_{\text{WGAN}}$ and subsequently optimising only the RL component ($\gamma = 0.0$). We fix a gradient penalty of $\lambda = 10$ and update the discriminator every five generator steps. For multi-property optimisation (QED, LogP, SA), we set weights of $w_1=0.4$, $w_2=0.3$, and $w_3=0.3$, respectively, following Tang et al \cite{tang2024molecular}. We assign zero reward to invalid molecules to enforce validity. For the QCBM, we  use a 16-qubit, 2-layer quantum circuit  and train it for 50 epochs with SPSA \cite{maryak1999efficient} using 1000 measurement shots per epoch. We fix  the circuit parameters after epoch 225 and we evaluate the model using 1000 samples per epoch.

\subsection{Evaluation Results}

Table ~\ref{tab:qcbm_metrics} summarises the performance of the QCA-MolGAN architecture under different reinforcement learning goals (QED, LogP, SA) and a MARL strategy. We evaluate chemical properties and metrics when the drug candidacy score is at a maximum, i.e when the generated samples are most likely to form a drug. Our QCBM-based prior achieves near-perfect chemical validity ($\geq \%99.7$) and novelty ($\geq \%99.4$) across invididual RL goals (QED, logP, SA) as well as under MARL, indicating that the generator reliably respects valence rules and explores outside the training set. Single-objective priors behave as intended: QCBM(QED) yields the best QED score of $0.607 \pm 0.015$, QCBM(SA) maximises synthesizability $0.946 \pm 0.064$, and QCBM(LogP) improves LogP relative to other single-objective runs $0.685 \pm 0.013$.

However, the MARL strategy, which jointly optimises multiple objectives, achieves the strongest macro average across properties (0.721), with top scores in SA and LogP at the expense of maximising the QED score. We observe that whilst training moves from single to multi-objective optimisation, there is a sharp drop-off in the uniqueness of samples falling to $4.5 \pm 0.6 \% $  under MARL. This is a clear sign that the GAN mechanism has undergone mode collapse onto a small subset of high-reward samples. This in turn, with the high diversity score $(1.000 \pm 0.001)$ implies that although collapsed onto a few high-scoring molecular modes each mode is mutually dissimilar. Overall, these results demonstrate that the QCBM prior in our QCA-MolGAN framework can be tailored to emphasise desired chemical properties.

\section{Conclusion}

In this work, we introduced QCA-MolGAN, a hybrid quantum–classical generative framework utilising a QCBM prior to guide molecular generation toward specific chemical properties. Our experiments on the QM9 subset showed that quantum priors optimised for individual properties (QED, LogP, SA) achieved targeted enhancements in their key chemical properties whilst also achieving high scores in validity and novelty in all cases. However, the MARL strategy further improved on almost all scores, most effectively balancing chemical property scores across generated molecules. To improve QCA-MolGAN, future work will focus on further enhancing multi-objective optimisation, potentially by developing a dynamic reward weighting scheme responsive to generation performance. Additionally, implementing an annealing schedule for the hyperparameter $\gamma$, initially favouring the WGAN objective and gradually transitioning dominance to the RL network, could improve stability of training and improve scores such as the uniqueness. To better understand our model performance, we will also benchmark our model on the full QM9 dataset and compare to the classical MLP variant.


\bibliography{refs}
\bibliographystyle{ieeetr}

\end{document}